# Can Predominant Credible Information Suppress Misinformation in Crises? Empirical Studies of Tweets Related to Prevention Measures during COVID-19

YAN WANG[1*], SHANGDE GAO[2], WENYU GAO[3]

[1*]Assistant Professor, Department of Urban and Regional Planning and Florida Institute for Built Environment Resilience, University of Florida, P.O. Box 115706, Gainesville, FL 32611, U.S. (*corresponding author*); E-mail: yanw@ufl.edu; ORCID: 0000-0002-3946-9418.

[2.] PhD Student, Department of Urban and Regional Planning and Florida Institute for Built Environment Resilience, College of Design, Construction and Planning, University of Florida, 1480 Inner Road, Gainesville, FL, 32601, U.S.; Email: gao.shangde@ufl.edu.

[3.] Postdoctoral Research Fellow, Harvard T.H. Chan School of Public Health and Department of Biostatistics, Harvard University, 655 Hutington Ave, Boston MA 02115, E-mail: wgao@hsph.harvard.edu; ORCID is 0000-0002-2128-9232.

**ABSTRACT**

During COVID-19, misinformation on social media affects people's adoption of appropriate prevention behaviors. It is urgent to suppress the misinformation to prevent negative public health consequences. Although an array of studies has proposed misinformation suppression strategies, few have investigated the role of predominant credible information during crises. None has examined its effect quantitatively using longitudinal social media data. Therefore, this research investigates the temporal correlations between credible information and misinformation, and whether predominant credible information can suppress misinformation for two prevention measures (i.e. topics), i.e. wearing masks and social distancing using tweets collected from February 15 to June 30, 2020. We trained Support Vector Machine classifiers to retrieve relevant tweets and classify tweets containing credible information and misinformation for each topic. Based on cross-correlation analyses of credible and misinformation time series for both topics, we find that the previously predominant credible information can lead to the decrease of misinformation (i.e. suppression) with a time lag. The research findings provide empirical evidence for suppressing misinformation with credible information in complex online environments and suggest practical strategies for future information management during crises and emergencies.

# 1 INTRODUCTION

Crisis communication plays a critical role in organizing effective responses and mitigating the impacts of crises (Clark-Ginsberg and Petrun Sayers, 2020). It can help people form the correct perceptions about prevention measures towards crisis events (Qiu and Chu, 2019) through disseminating credible information regarding the assessment and mitigation of crisis events (Gilk, 2007) as well as guidance about correct response measures (Utz, Schultz, and Glocka, 2013). Social media platforms facilitate the process of crisis communication by allowing people to seek, interpret, and disseminate information during crisis events (Silver and Andrey, 2019). During COVID-19, social media has been ignited with a diversity of information. The increasing rate of detected incidents along with massive, related dialog has triggered divergent reactions and interactions across stakeholders at various levels (Shimizu, K. 2020; Wang et al. 2021). Specifically, under the social distancing policy, more people have turned to social media for support (Nabity-Grover, et al. 2020). However, the credibility of social media information is worrisome. Misinformation, i.e. inaccurate or misleading information (Vosoughi et al., 2018) spreads widely and quickly (Depoux et al., 2020; Pulido et al., 2020). This posed severe challenges to the public especially during public health crises such as COVID-19. For example, Kouzy et al. (2020) found that after the worldwide outbreak of coronavirus disease in 2019 (COVID-19), 24.8% of tweets about COVID-19 contained misinformation. Unlike credible information, which contains positive attitudes towards the correct prevention measures of the crises (Castillo, Mendoza, and Poblete, 2011), most misinformation contains negative attitudes towards the correct measures and produces misperceptions about disease prevention (van der Meer and Jin, 2020).

Misinformation during public health crises is harmful because it misdirects people's response behaviors while the effectiveness of intervention policies depends heavily on individuals' response behaviors. For example, the spread of coronavirus can be controlled by individual-level prevention strategies, such as wearing facemasks (Feng et al., 2020) and social distancing (Lewnard and Lo, 2020). Individuals' crisis response behaviors can be significantly affected by information obtained from the Internet and social media (Swire-Thompson and Lazer, 2020). However, some factors, such as recommendation algorithms and bots, have made misinformation widely propagate in the digital environments (Zhang and Ghobani, 2020, Orabi et al., 2020). Individuals misled by such misinformation may avoid following the correct recommendations and put their health at high risk (Earnshaw and Katz, 2020). For example, a widespread coronavirus treatment of "injecting disinfectant" caused 30 poisoning cases in New York City within 18 hours (Slotkin, 2020). Additionally, misinformation specifically has a heavy impact on vulnerable groups during the COVID-19 pandemic: mistrust and lack of access to credible information sources made the vulnerable groups easily to be affected by misinformation (Clark-Ginsberg and Petrun Sayers, 2020). Because of the vast spread and negative health impacts of misinformation, it is urgent to formulate effective strategies to suppress misinformation on social media platforms.

Previous literature has proposed several strategies for combating crisis-related misinformation on social media, including checking information authenticity (Safieddine, et al. 2016), controlling bot accounts (Shao et al., 2018), tracking sources of misinformation (Jang et al., 2018), identifying misinformation topics (Vicario et al., 2019), broadening exposure to diverse views (Wang and Song, 2020), and providing news and science literacy education, such as guidelines of social media usage in crisis events (Kaufhold et al., 2019; Trethewey, 2020; Tully, et al. 2020). The first five strategies can be implemented by social media companies, while the last one puts the onus on the public and authoritative agencies. Specifically, in the domain of public health, fact-checking (conducted by social media platforms and experts) and literacy education (Walter et al., 2020) have been used as the main strategies to suppress health misinformation on social media. However, the effectiveness of fact-checking and bot control has been limited to suppressing pre-known misinformation. Detecting misinformation and checking facts is not feasible with large datasets (Shao et al., 2018) and cannot limit the production or sharing of posts containing undetected misinformation. In addition, controlling bot accounts cannot mitigate the misinformation generated and shared by human accounts.

In comparison with the detection-based "reactive" strategies, literacy education (e.g. news and information literacy) has a greater potential to suppress misinformation "proactively" (Tully, Vraga, and Bode, 2020). Literacy education reduces the public's ignorance and misconceptions on specific topics, such as climate change (Cook, et al. 2014), and helps individuals correctly judge the truthfulness of information (Kahne and Bowyer, 2017). For public health, literacy education helps people form correct perceptions of disease conditions and prevention measures. The effectiveness of literacy education has been notable, and experiments have shown that the provision of accurate information made about around 20% of the experiment participants change their misperceptions of the research topics (Vraga, Bode, and Tully, 2020). On social media platforms, literacy education has been applied by disseminating credible information about crisis events and providing correct strategies for crisis prevention (Almaliki, 2019). In the public health domain, the strategy of disseminating credible information has also been used to suppress misinformation, especially in vaccination promotion and COVID-19 prevention (Danielson, Marcus, and Boyle, 2019; Chen, Lerman, and Ferrara, 2020). For example, during COVID-19, social media posts containing credible information about COVID-19 prevention and published by authoritative information sources such as the Centers for Disease Control and Prevention (CDC) and the World Health Organization (WHO) (Chen, et al. 2020) were spread widely.

However, little research has investigated the temporal correlation between misinformation and credible information during crises empirically, and the existing research remains insufficient on whether predominant credible information can effectively suppress misinformation on social media platforms. None has used longitudinal social media data to investigate the temporal relationship between misinformation and credible information quantitatively. It is unclear how effectively the previously predominant credible information (e.g. increased number or proportion) can reduce the overall volume and proportion of misinformation on social media during crises. Considering the existing research gaps and the urgency of suppressing social-media misinformation about COVID-19 prevention, this manuscript has two primary questions.

> **#1**: What is the temporal relation between the daily volume/proportion of tweets that contained credible information and misinformation for individual topics of prevention measures?
>
> **#2**: Can previous predominant credible information suppress misinformation on Twitter?

We chose two topics, i.e. "wearing masks" and "social distancing," for detailed empirical investigations (Feng et al., 2020; Lewnard and Lo, 2020) due to the potential negative influence of their misinformation on individuals' prevention behaviors during COVID-19. These prevention measures affect people's healthy mobility and interactions with their built environments, and various types of misinformation have been found on Twitter that might hinder people from following these measures (Krause et al., 2020). Particularly for the two public health topics, our classification criteria are built based on the potential public-health consequences of the two information categories. We regarded tweets as credible if they supported the two critical prevention measures and affirmed the negative consequences of not following them, and as misinformation if they opposed these measures (van der Meer and Jin, 2020; Wilson and Starbird, 2020). If people opposed verified measures for COVID-19 prevention, they tended to behave inappropriately in response to the COVID-19 pandemic or to share such attitudes on social media platforms, and their health status would be highly risky (Earnshaw and Katz, 2020). "Wearing masks" and "social distancing" refer to effective measures for COVID-19 prevention, and their effectiveness has been verified by medical experiments (Feng et al., 2020; Lewnard and Lo, 2020). Acting on misperceptions of these methods, such as not wearing a mask or socially distancing in public places, would accelerate the spread of coronavirus; one experiment showed that a lack of appropriate prevention measures would nearly double the number of infections over a situation with proper prevention measures (Lewnard and Lo, 2020). Based on the medical evidence above, we regarded tweets containing the negative attitudes towards these measures as misinformation. We utilized key-expressions and Support Vector Machine (SVM) to extract relevant tweets from the collected data and categorized them into those containing (a) credible information and (b) misinformation under each topic. We generated the series about the daily volume and proportion of these two information categories, then conducted cross-correlations between the time series of two information categories. The research findings can provide strategies for combating social media misinformation during future public health crises and other extreme events.

## 2 DATA COLLECTION AND METHODS
### 2.1 Case Description and Data Collection

This research focused on the misinformation about coronavirus prevention on Twitter and evaluated the influence of credible information on the spread of misinformation in the U.S. The study period covers 136 days, from February 15 to June 30, 2020. We chose this period because the number of U.S. cases of coronavirus proliferated after the Diamond Princess Event (CDC, 2020) and surpassed three million during the second wave of the pandemic (Dong, et al. 2020). During this time, a large volume of misinformation spread widely (Hernández-García and Giménez-Júlvez, 2020), which caused an infodemic and potentially sped up the virus transmission. The misinformation about the prevention measures, such as recommendations not to wear masks, to ignore social distancing, and to engage in risky behaviors (Pennycook et al., 2020), hinders the use of proper prevention measures. Meanwhile, however, credible information was also disseminated to inform individuals of the proper response measures and to suppress the misinformation.

Over the four-and-half months, we collected tweets with keywords "coronavirus" and "covid" using an open Twitter streaming API (Twitter, 2020b). We focused on English tweets, as they represented the majority of Twitter users in the U.S., and we retrieved 28,573,952 English tweets from the raw data. In addition, because the streaming API could not retrieve the full texts of tweets (most were truncated), we used Hydrator (Documenting the Now, 2020) to extract the full text of each tweet before further text mining. With the basic datasets, we then conducted machine-learning-based analyses in three steps: (i) retrieving relevant tweets, (ii) classifying tweets as containing misinformation and as containing credible information, and (iii) investigating the cross-correlation between time series of the two information categories (see Figure 1).

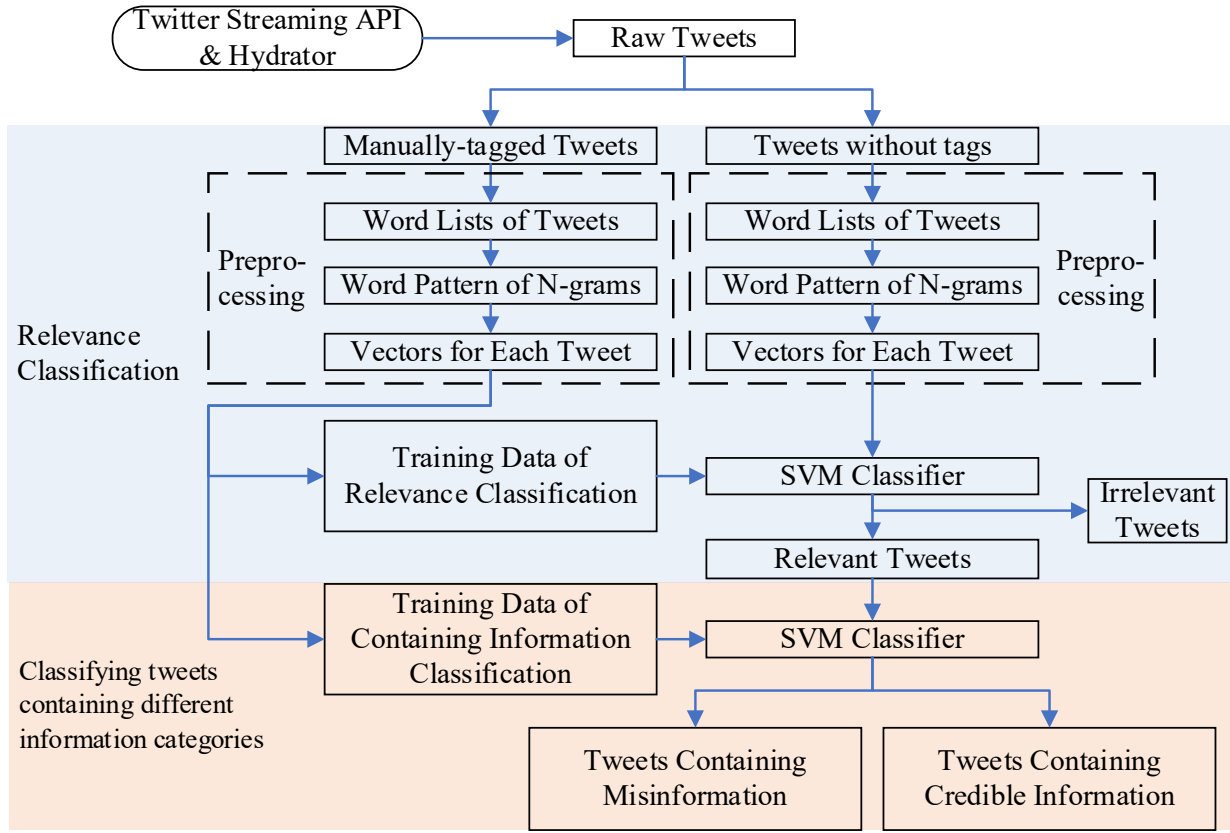

**Figure 1**. Schematic process of tweets analysis

**2.2 Retrieving Relevant Tweets for the Two Preventative Measures**

We conducted two steps to extract tweets that are relevant to each topic, including initial keyword-based filtering and supervised classification using SVM. We define "relevant tweets" as (i) tweets that directly expressed opinions on the three topics, such as "wearing masks is useful"; (ii) tweets involving suggestions, policies, or opinions in a certain area or (iii) tweets that endorsed suggestions, policies, and opinions about any of the preventative measures, such as "Dr. Fauci did not recommend wearing masks". First, we used key-expression filtering to retrieve tweets containing the keywords and expressions for each topic (see Table 1). To generate the final keyword list, we first collected key expressions about the topics from the websites of the U.S. CDC (2020) and the WHO (2020). Then we used both the keywords (e.g., "wearing masks") and their expression patterns to collect the potentially relevant tweets. For example, we used the pattern "*'mask' + 'second waves study'*" to retrieve tweets containing both phrases. Using such patterns, we could collect tweets that did not use the specific format of our keywords but still contained relevant content. For example, both "*masks can effectively protect others*" and "*to protect others, masks are necessary*" contain the pattern "*'mask' + 'protect others'*", but the forms of the key expression are not the same, and we cannot retrieve tweets containing such contents using a single specific keyword. Because of the disadvantages of keyword-based filtering, we also used patterns of keywords (i.e., key-expressions) to filter out tweets that were potentially relevant to the topics. Using this list of key expressions and patterns, we retrieved three thousand sample tweets from the API-collected data and enriched this list based on tweet texts.

**Table 1**: Final keywords list for each topic and number of filtered tweets

| Topics | Key-expressions | Data volume |
|---|---|---|
| Mask | 'wear a mask', 'wearing a mask', 'wearing face mask', 'wear face masks', 'wear your mask', 'mask-wearing could prevent', 'mask' + 'second waves study', 'mask in public', 'mask protects you', 'mask' + 'please please please', 'mask' + 'prevent the spread', 'mask' + 'prevent you from', 'mask' + 'slow the spread', 'use of facemask', 'mask won't help', 'masks at all times', 'masks are useless', 'mask is useless', 'face coverings', 'facemask use', 'healthy people', 'masks can', 'N95 masks', 'prevent COVID-19', 'please wear', 'mask' + 'protect others', 'mask' + 'protect themselves', 'mask' + 'protect yourself', 'mask' + 'protects you', 'wear mask', 'wearing masks', 'need mask', 'wore mask', 'no mask', 'mask' + 'effectiveness', 'mask' + 'efficiency', 'mask' + 'compulsory', 'WearAMask', 'mask' + 'reduce onward transmission'. | 667,761 |
| Social Distancing | 'social distancing', '2 arms', '6 feet', '6-foot distance', 'avoid crowded places', 'avoid crowds', 'avoid gathering', 'avoid hugging', 'avoid kissing', 'avoid pooled rides', 'close contact', 'common areas', 'create space between others', 'face-to-face contact', 'increase space between individuals', 'keep a safe space', 'keep distance', 'keep space', 'limit contact', 'limit errands', 'physical distance', 'physical guide', 'safe social activities', 'social distance', 'stay apart', 'stay distanced', 'physical distancing', 'around others'. | 101,113 |

However, key expressions could still retrieve irrelevant tweets. For example, the tweet "Coronavirus: 3M to Produce 35,000,000 Respirator Masks a Month in the U.S." contains "coronavirus" and "mask", but it is about mask production instead of behaviors of wearing masks, so we regard it as irrelevant. This tweet contains keywords about social distancing (i.e. "social distancing") but the three rules about relevance classification deem it irrelevant because it does not contain opinions or suggestions about social distancing.

To overcome the limitations of key-expressions in retrieving relevant tweets, considering the high-level performance of SVM in text classification, of which the accuracy was higher than 90% (Liu et al., 2013; Gopi et al., 2020), we conducted the second step of relevance classification using an SVM-based classifier. The training datasets were randomly extracted from the raw dataset over the whole study period. Sentences of tweets in the training datasets and case study dataset were tokenized to unigrams, bigrams, and trigrams using the NLTK Tokenizer, and then vectorized using the TF-IDF algorithm because TF-IDF can reflect how relevant a given word is in a particular document (Ramos. 2003). We used the vectors of the training data to train the SVM-based classifier and then used the classifier to label the tweets of the case study dataset. The training outcome of the SVM-based classifier is shown in Table 2. The outcomes of the relevance classification were two datasets that contained relevant tweets and irrelevant tweets. The relevant tweets were used for further information classification.

Table 2. Training and testing dataset's sizes and classification performance for relevant tweets

|  | Wearing mask | Social distancing |
|---|---|---|
| Training Dataset | 1,192 | 2,099 |
| Testing Dataset | 300 | 300 |
| Accuracy | 0.8833 | 0.9133 |
| Precision | 0.9099 | 0.9462 |
| Recall | 0.9380 | 0.9535 |

**2.3 Classifying Tweets Containing Credible Information and Misinformation**

After manually annotating tweets containing the two categories of information based on the criteria described in Introduction, we trained the SVM-based classifier to classify tweets under each topic over the four and half months, so that we could analyze the relationship between the volumes of tweets containing credible information and misinformation from a temporal perspective. The training outcome of the SVM-based classifier is shown in Table 3.

Table 3. Training and testing datasets and classification performance for information categories

|  | Wearing mask | Social distancing |
|---|---|---|
| Training Dataset | 1,684 | 941 |
| Testing Dataset | 300 | 300 |
| Accuracy | 0.9641 | 0.8333 |
| Precision | 0.9447 | 0.8507 |
| Recall | 0.9305 | 0.9495 |

**2.4 Cross-Correlation Analysis of Credible and Mis-information Time Series**

Time-series analyses have been widely used in analyzing data and information mined from social media platforms (e.g. Wang and Taylor 2018). We employed a cross-correlation analysis of two time series to identify lags ($h$) of the predominant daily volume/proportion of credible information ($c_{t+h}/cp_{t+h}$) that might be useful predictors of daily volume/proportion of misinformation ($m_t/mp_t$) for tweets relevant to "wearing masks" and "social distancing" topics separately. For example, when one or more $c_{t+h}$, with $h$ negative, are predictors of $m_t$, it is sometimes said that $c$ leads $m$; when one or more, $c_{t+h}$ with $h$ positive, are predictors of $m_t$, it is sometimes said that $c$ lags $m$.

The cross-correlation analysis is performed based on the plot of cross-correlation function (CCF) between the time series of credible information and misinformation (i.e. daily tweet count and daily proportion) for each topic. Values of the x-axis of the peaks in CCF plots indicate potential significant time lags on the predictor (i.e. credible information). Before running CCF, a pre-whitening procedure using an Autoregressive Integrated Moving Average (ARIMA; Box et al., 2015) model is used to remove the common trends of time series of two information categories and to help better interpret the CCF. The final model is constructed with the final chosen lags based on the CCF plot and the ARIMA model.

To perform pre-whitening, we fit the ARIMA model to the predictor ($c_t/cp_t$) and use the fitted model structure to filter out the response ($m_t/mp_t$). The ARIMA also requires stationarity (i.e. the mean and variance do not change over time). We conducted the Augmented Dickey–Fuller (ADF) Test (Said and Dickey 1984) and the analysis is performed using the *adf.test* function from the "tseries" R package (Trapletti and Hornik, 2020). If $p-value$ of the ADF test is less than 0.05, the time series is stationary; if $p-value$ of ADF test is equal to or larger than 0.05, the time series is not stationary. As the function cannot pass missing values, we imputed missing data using *Kalman* smoothing (Harvey, 1990; Bishop and Welch, 1995; Grewal et al., 2020), a nonparametric method without model assumptions. This process employed a *na_kalman* function from package "imputeTS" (Moritz and Bartz-Beielstein, 2017). If the time series is stationary, the ARIMA model is fitted using the *sarima* function from the "astsa" package (Stoffer, 2020).

To select the best ARIMA and final cross-correlation models, we start the fitting with all the candidate time lags, then use backward selection. The ARIMA model selection criteria are based on Akaike information criterion (AIC; Akaike, 1998), and Bayesian information criterion (BIC; Schwarz, 1978) (For AIC and BIC, the smaller the better), and ensure the residuals to be independent (ACF around zero) and random (Ljung-Box test with $p-value > 0.05$) (Ljung and Box, 1978). The final cross-correlation model is linear so our selection is based on adjusted R-squared (Draper and Smith, 1998), ensuring the residuals to be independent (ACF around zero) in model validation. Essential time series plots including CCF, autoregressive function (ACF), and partial autoregressive function (pACF), were made using the R built-in package 'stats'. All the statistical analyses were performed in R (R Core Team, 2020).

## 3 RESULTS

### 3.1 Tweets containing Credible Information and Misinformation

We utilized the key-expressions (Table 1) and SVM-based classifiers to retrieve the relevant tweets of case topics (i.e. wearing masks and social distancing) from the tweets collected from February 15 to June 30 (using methods described in Section 2.2 and 2.3). We have 12 days with missing data from April 21 to 28 and June 6 to 9 due to the tropical-storm-incurred power outages in Florida and computer resetting, which has a very minor impact on the following analyses based on 4.5-month data. The changes in the daily volume of tweets that contain misinformation and credible information are plotted in Figure 1.

Based on the daily data volume of the classified tweets (Figure 1), we find that tweets relevant to "wearing masks" kept growing over periods of (a) February 15 to 29 and (b) April 4 to May 30, potentially caused by the increasing public attention on the reasonability and implementation of wearing masks. The second period of growth might be intensified by the event of George Floyd on May 25 (Dave et al., 2020), when people protested for the policemen's violence in Minneapolis. Additionally, as the number of U.S. cases surpassed 100,000 on May 28 (Dong, Du, and Gardner, 2020), CDC highly recommended individuals wearing masks in public places, which could contribute to the increased discussions as well. In comparison, the number of "social distancing" tweets did not change drastically and grew from February 15 to June 6 steadily, then decreased gradually. Based on the health literature (e.g. Lewnard and Lo, 2020), social distancing was proved as an effective strategy and continuously promoted by public health agencies, and the discussion of social distancing on Twitter was growing from middle February to early June. The prevention measure was promoted by the persistent recommendation of the related public health policies (Chui et al., 2020), but the popularity level of the discussion was not as high as "wearing masks". To understand the proportion of credible information and misinformation in the two datasets containing topic-relevant tweets, we also calculated the daily proportion for each topic (see Figure 2).

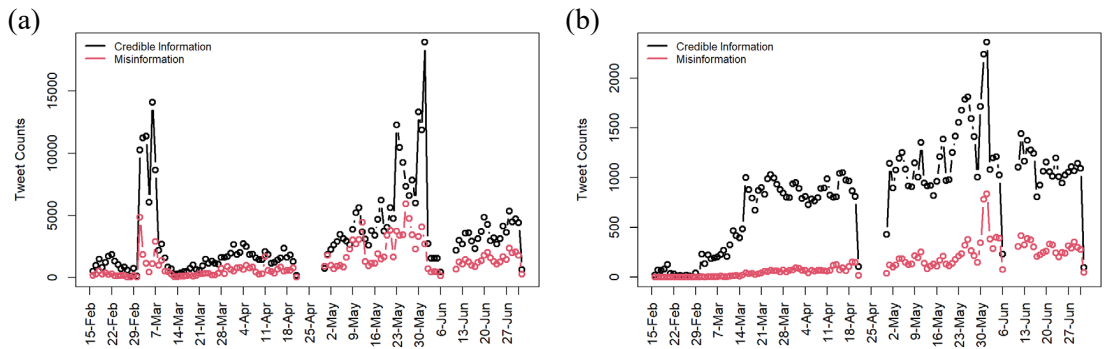

**Figure 1.** Daily number of tweets containing misinformation and credible information (a: wearing masks; b: social distancing)

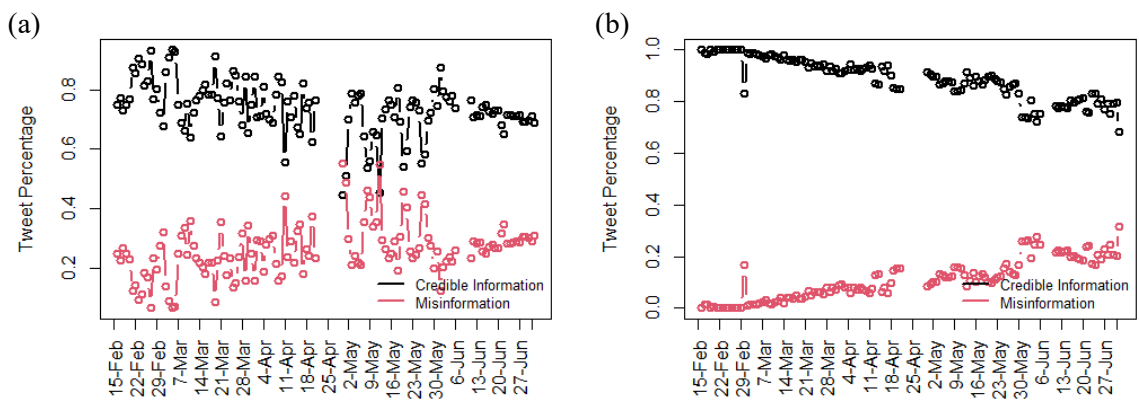

**Figure 2.** Daily percentage of tweets containing misinformation and credible information (a: wearing masks; b: social distancing)

### 3.2 Cross-Correlation between "Wearing Masks" Credible and Misinformation Time Series

To explore the relation between misinformation and credible information for the "wearing masks" topic over time, we employed cross-correlation analysis of time series in both original-number and percentage scales, including (a) daily tweet number containing misinformation ($m_t$) and credible information ($c_t$); and (b) daily proportion of misinformation ($mp_t$) and credible information ($cp_t$). The initial CCF plots (SM Figure 1) for time series in both scales showed unclear peaks of time lags, so pre-whitening is conducted.

Specifically, for "wearing mask", the ADF test ($p-value = 0.05384 > 0.05$) indicates that the time series ($c_t$) is not stationary (Fuller, 2009), so we took the first-order difference of predictor between the daily values of adjacent dates ($c_t - c_{t-1}$). Then the ADF test ($p-value < 0.01$) shows the time series of the predictor's first-order difference is stationary. Thus, we considered integration with Order 1. Our final ARIMA model chose AR with Order 6, because of the integration of order 1, we considered time lag ($t = 1, 2, 3, 3, 6, 7$) eliminating $t = 5$ after model selection (see Methods). The final cross-correlation model is listed in Equation 1 with detailed coefficients and significance levels in SM Table 1 and satisfactory ACF and pACF tests for model validation in SM Figure 3. The adjusted $R^2$ for the final model is 0.7041 with $p-value < 2.2e-16$.

For $cp_t$ and $mp_t$, after imputing missing values, the ADF test has a $p-value < 0.01$, indicating stationary. The pre-whitened CCF plot (SM Fig 1b) indicated potential important time lags ($h$) at 0, -9 and -12. Based on the fitting outcomes of ARIMA model, time lags at -1, -19, and -20 were also considered. Notably, time lag at 0 is omitted due to the collinearity with the response variable. Thus, we chose the final model based on adjusted $R^2$ and ACF tests (SM Fig 4) and the model is listed in Equation 2 with detailed coefficients and significance in SM Table 2 and satisfactory ACF and pACF tests for model validation (SM Figure 4). For the final model, the adjusted $R^2$ is 0.2273, and the $p-value = 4.156e-05$.

Based on the final fitted cross-correlation models (Equation 1 and 2) and the significance of coefficients in SM Table 1 and 2, we find evidence that predominant credible information (i.e. tweet number and percentage) leads the decrease of misinformation significantly when lag ($h$) is – 1 (one day). However, we also find that misinformation tweets from the previous day and the credible tweets from the same day have a positive significant correlation with the number of tweets containing misinformation; the number of misinformation tweets can also negatively impact the number of credible tweets in the future with a time lag at 10. Additionally, the number of tweets containing misinformation is also positively related to the time ($t$) significantly. For the daily percentage of misinformation tweets, previous dominant credible information in percentages with a time lag at -1, -9, -19 can all significantly decrease the percentage of misinformation for wearing masks tweets.

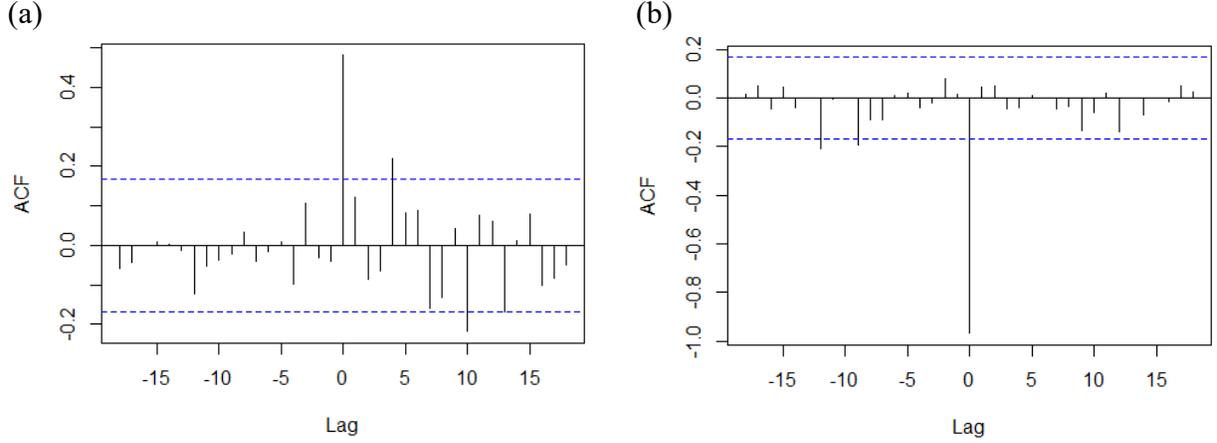

**Figure 3.** Cross-Correlation of $M_t$ and $C_t$ of wearing masks after pre-whitening based on different lags (a: original scale; b: percentage scale)

$$\begin{cases} m_t = -384.09160 + 0.27772m_{t-1} - 0.16286m_{t-7} + 0.13063c_t - 0.08387c_{t-1} + 0.11028c_{t-3} \\ \qquad\qquad + 0.12101c_{t+4} - 0.02947c_{t+10} + 11.49660t + \epsilon_t \\ \epsilon_t \sim N(0, \sigma^2) \end{cases} \quad (1)$$

$$\begin{cases} mp_t = 0.71031 - 0.24856cp_{t-1} - 0.19105cp_{t-9} - 0.16108cp_{t-19} + \epsilon_t \\ \epsilon_t \sim N(0, \sigma^2) \end{cases} \quad (2)$$

### 3.3 Cross-Correlation between "Social Distancing" Credible and Misinformation Time Series

Similarly, we conducted cross-correlation analyses and ADF test for the time series of "social distancing" tweets containing misinformation and credible information. The CCF plots for the two scales (daily number and proportion) shown in SM Figure 2 indicates that pre-whitening is necessary. The CCF plots after the pre-whitening process are in Figure 4.

For original tweet number under each information categories ($m_t$ and $c_t$), the ADF test on $c_t$ has a $p-value$ of 0.4388 (non-stationary). After taking the first-order difference of the predictor, the $p-value$ of ADF test is smaller than 0.01. Thus, an order (1) integration is considered. The $ARIMA(12, 1, 0)$ model omitting order 1, 8, 9, and 10 is used. The preliminary model based on adjusted $R^2$ is listed in Equation 3. Coefficients and significance in SM Table 3, the adjusted $R^2$ is 0.859, and the model $p-value < 2.2e-16$. However, the residuals also have autoregression (based on the ACF and pACF tests in SM Figure 5), so we fitted the cross-correlation model considering the autoregressive residuals ($w_t$) simultaneously. The final model is listed in Equation 4 and the value of $AIC$ is 6.418127, $AICc$ is 6.446242, and $BIC$ is 6.686266. Detailed coefficients and significance can be found in SM Table 4 with satisfactory ACF and pACF tests for model validation (SM Figure 5).

For the daily proportion of each information category ($mp_t$ and $cp_t$), the ADF test has $p-value$ of 0.2138 after imputation, and $p-value < 0.01$ after taking first-order difference (indicating stationarity). Thus, order (1) integration is considered. An $ARIMA(19,1,0)$ keeping orders at 1 to 4 and 19 is chosen. The pre-whitened CCF plot is shown in Figure 4. Lags to be considered include -1, -2, -3, -4, -7, -19, 11, and 16. The final model considering both adjusted $R^2$ and $p-values$ is listed in Equation 5 with coefficients and significance in SM Table 5 and satisfactory ACF and pACF tests for model validation (SM Figure 6). The adjusted $R^2$ is 0.8639, and the model $p-value < 2.2e-16$.

Based on the final fitted cross-correlation models (Equation 4 and 5) and the significance of coefficients in SM Table 4 and 5, we find evidence that for the topic of "social distancing", predominant credible information (i.e. tweet number and proportion) leads the decrease of misinformation significantly, when lag ($h$) is – 3, -11, -12 for the number of tweets and -1 for the proportion of tweets. However, we find that the credible tweets from the same day also have a positive significant correlation with the number of tweets containing misinformation; the number of misinformation tweets can also negatively impact the number of credible tweets in the future with a time lag of 2 and 3 days. The percentage and number of tweets containing misinformation are also positively related to the time (t) significantly.

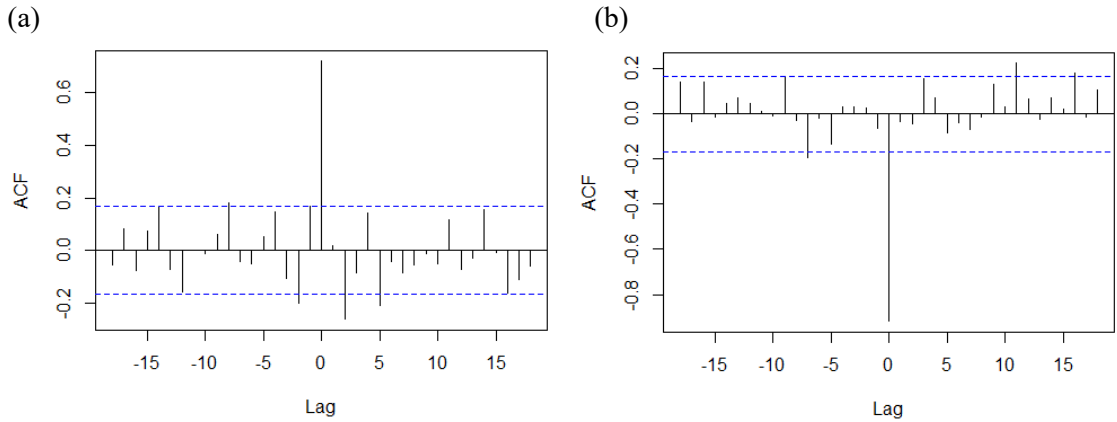

**Figure 4.** Cross-Correlation of $M_t$ and $C_t$ of social distancing after pre-whitening based on different lags (a: original scale; b: percentage scale)

$$m_t = -65.98106 + 0.24143c_t - 0.07276c_{t-3} + 0.07202c_{t-5} + 0.05695c_{t-6} + 0.04156c_{t-8} \\ -0.08628c_{t-11} - 0.08938c_{t-12} - 0.09473c_{t+2} - 0.07784c_{t+5} + 2.92643t + w_t \quad (3)$$

$$\begin{cases} m_t = -60.2682 + 0.1783c_t - 0.0452c_{t-3} + 0.0941c_{t-5} + 0.0583c_{t-6} + 0.0364c_{t-8} \\ \quad -0.0717c_{t-11} - 0.0644c_{t-12} - 0.1365c_{t+2} - 0.0506c_{t+5} + 2.9318t + w_t \\ w_t = 0.4294w_{t-1} + \epsilon_t + 0.7636\epsilon_{t-1} \\ \epsilon_t \sim N(0, \sigma^2) \end{cases} \quad (4)$$

$$\begin{cases} mp_t = 0.4943 - 0.4990cp_{t-1} + 0.0008941t + \epsilon_t \\ \epsilon_t \sim N(0, \sigma^2) \end{cases} \quad (5)$$

**4 DISCUSSION AND CONCLUSION**

**4.2 Findings and Contributions**

COVID-19, the worldwide drastic pandemic, has ignited online platforms and caused an "infodemic" on various channels of crisis communication; misinformation about prevention measures of coronavirus also spreads widely and have affected the adoption of proper prevention measures. Although studies (e.g. Wang, et al. 2020) have found that effective risk and crisis communication with credible information can positively impact the performance of public health campaigns and government agencies have also disseminated credible message on social media platforms actively, the temporal relation and the potential suppression effects of credible information on misinformation have not been investigated in detail empirically.

This research analyzed a big amount of longitudinal social media data using supervised machine learning methods and cross-correlation analyses of time series. It quantitatively investigated the temporal cross-correlation between credible information and misinformation and whether predominant credible information can suppress misinformation on Twitter. Our analyses found evidence about the suppression effects of previously predominant credible information on misinformation for the two preventive-measure topics on Twitter. Specifically, in tweets relevant to both topics of "wearing masks" and "social distancing", we found that the increasing percentage of credible information from the previous day led to a decrease in the percentage of misinformation significantly. The increasing number of tweets containing credible information from a previous day led to a decrease in the number of tweets containing misinformation significantly, while the significant time tags ($h$) for the two topics varied. In addition to the "suppression" effect of credible information (in scales of number and percentage) on misinformation, we also found that; (a) the number of misinformation-relevant tweets increased significantly over time for both topics; (b) the number of credible tweets from the same day also had a positive significant correlation with the number of misinformation tweets, and (c) the number of misinformation tweets also had significant correlations with the number of credible tweets in future days but the effects varied when the time lags were different.

This research advances the existing knowledge body of crisis communication and misinformation research, especially for studies focused on public health crises. Although spreading credible information has the potential to reduce the misinformation on social media platforms (Jin et al., 2020; Iosifidis and Nicoli, 2020), little research has found empirical evidence of predominant credible information's role in suppressing misinformation. To the best of our knowledge, none has quantified the suppression effect of credible information over time. To provide empirical and quantitative evidence of the suppression effect of credible information, we analyzed real-world social media posts (tweets) about the COVID-19 pandemic. Compared to the survey outcomes of previous research, this longitudinal dataset reflects the attitude change of general Twitter users towards COVID-19 prevention measures in real-world situations rather than in experimental scenarios. The research findings can guide public health authorities, emergency responders, and other crisis managers to actively disseminate and endorse credible information in online platforms in order to suppress misinformation increase aggregately over time. By developing evidence-based strategies, crisis managers can inform the public of appropriate prevention measures for COVID-19 as well as the damages caused by ineffective prevention behaviors more effectively to achieve crisis communication goals. This research also provided insights into methods for studying different categories of information during crises on social media, as we mined and revealed the temporal patterns of both credible information and misinformation on Twitter during COVID-19.

### 4.2 Limitations and Future Work

There are a few potential limitations of this study and opportunities for future research. First, it focused on English tweets collected by a keyword-based Twitter Streaming API. Future work might use accurate translation algorithms to process tweets in other languages before conducting English-based natural language processing. Data from other social networking platforms could also be considered if they become available. Second, the existing supervised machine learning methods, including SVM, cannot achieve 100% accuracy when classifying data. We have put considerable effort into raising the classifier's accuracy to the level between 85% and 95%, such as increasing the volume of training data, comparing classification algorithms, and manually annotating the training data, and overall, our final classifiers outperformed existing classifier used in similar tasks (e.g. Yao & Wang 2020). With further development of text mining techniques, researchers could use more advanced AI techniques to classify tweets containing different information categories and reveal the real-world situation of information dissemination more accurately. Third, we classified the tweets containing misinformation and credible information mainly based on users' attitudes towards the correct prevention measures of COVID-19, but such criteria may not apply to crises that do not affect public health. Future research can extend the investigation of relations between credible information and misinformation in other types of extreme events and crises, including natural hazards and political crises.

**Acknowledgments**

This material is based upon work supported by the National Science Foundation under Grant No. 2028012. Any opinions, findings, and conclusions or recommendations expressed in this material are those of the authors and do not necessarily reflect the views of the National Science Foundation.

# Supplementary Material

**CCF Plots**

**SM Figure 1.** Cross-correlogram (CCF Plots) of the time series of two information categories based on different lags for Wearing Masks. (a: CCF of original scale; b: CCF of percentage scale)

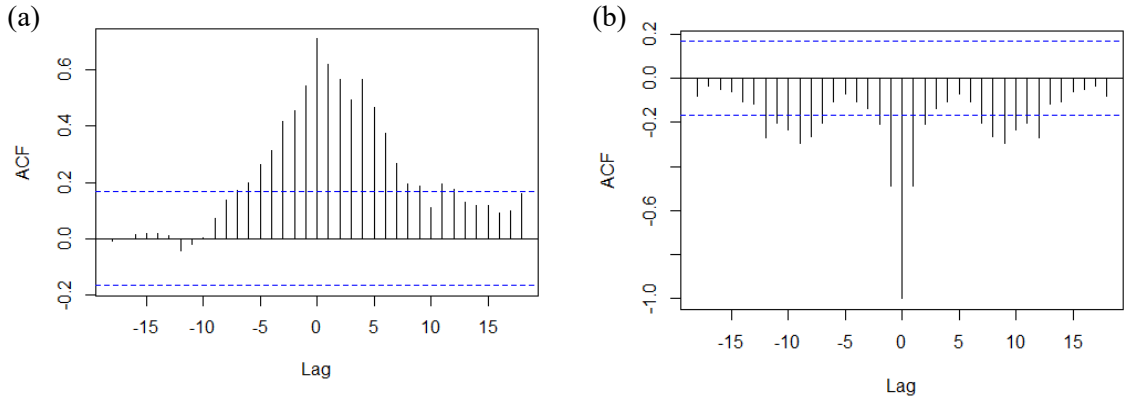

**SM Figure 2.** Cross-correlogram (CCF Plots) of time series of two information categories based on different lags for Social Distancing. (a: CCF of original scale; b: CCF of percentage scale)

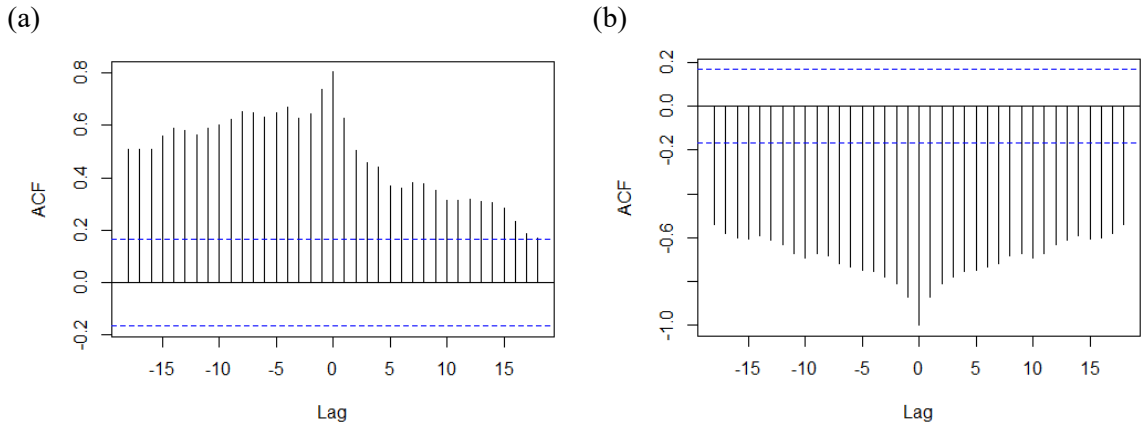

**CCF Tables**

**SM Table 1.** Coefficients of Final Cross-Correlation Model for Wearing Mask (Daily Number)

|  | $Estimate$ | $Std.Error$ | $t\ value$ | $Pr(>|t|)$ | Significance |
|---|---|---|---|---|---|
| $(Intercept)$ | -384.09160 | 182.20216 | -2.108 | 0.038715 | * |
| $m_{mask,t-1}$ | 0.27772 | 0.10623 | 2.614 | 0.010999 | * |
| $m_{mask,t-7}$ | -0.16286 | 0.09437 | -1.726 | 0.088937 | . |
| $c_{mask,t}$ | 0.13063 | 0.03363 | 3.885 | 0.000235 | *** |
| $c_{mask,t-1}$ | -0.08387 | 0.04075 | -2.058 | 0.043422 | * |
| $c_{mask,t-3}$ | 0.11028 | 0.03736 | 2.952 | 0.004334 | ** |
| $c_{mask,t+4}$ | 0.12101 | 0.02697 | 4.488 | 2.85e-05 | *** |
| $c_{mask,t+10}$ | -0.02947 | 0.02454 | -1.201 | 0.233924 |  |
| $t$ | 11.49660 | 3.24644 | 3.541 | 0.000724 | *** |

'***', '**', '*' and '.' describe significance levels at 0.001, 0.01, 0.05, and 0.1 respectively.

**SM Table 2.** Coefficients of Final Cross-Correlation Model for Wearing Mask (Daily Proportion)

|  | $Estimate$ | $Std.Error$ | $t\ value$ | $Pr(>|t|)$ | Significance |
|---|---|---|---|---|---|
| $(Intercept)$ | 0.71031 | 0.08747 | 8.120 | 5.96e-12 | *** |
| $cp_{mask,t-1}$ | -0.24856 | 0.09661 | -2.573 | 0.0120 | * |
| $cp_{mask,t-9}$ | -0.19105 | 0.08621 | -2.216 | 0.0296 | * |
| $cp_{mask,t-19}$ | -0.16108 | 0.07880 | -2.044 | 0.0444 | * |

'***', '**', '*' and '.' describe significance levels at 0.001, 0.01, 0.05, and 0.1 respectively.

**SM Table 3.** Coefficients of Preliminary Cross-Correlation Model for Social Distancing (Daily Number)

|  | $Estimate$ | $Std.Error$ | $t\ value$ | $Pr(>|t|)$ | Significance |
|---|---|---|---|---|---|
| $(Intercept)$ | -65.98106 | 15.89398 | -4.151 | 9.96e-05 | *** |
| $c_{dist,t}$ | 0.24143 | 0.03400 | 7.100 | 1.24e-09 | *** |
| $c_{dist,t-3}$ | -0.07276 | 0.04431 | -1.642 | 0.105494 |  |
| $c_{dist,t-5}$ | 0.07202 | 0.05582 | 1.290 | 0.201611 |  |
| $c_{dist,t-6}$ | 0.05695 | 0.05574 | 1.022 | 0.310741 |  |
| $c_{dist,t-8}$ | 0.04156 | 0.04521 | 0.919 | 0.361409 |  |
| $c_{dist,t-11}$ | -0.08628 | 0.04814 | -1.792 | 0.077800 | . |
| $c_{dist,t-12}$ | -0.08938 | 0.04386 | -2.038 | 0.045714 | * |
| $c_{dist,t+2}$ | -0.09473 | 0.03043 | -3.113 | 0.002771 | ** |
| $c_{dist,t+5}$ | -0.07784 | 0.02171 | -3.585 | 0.000652 | *** |
| $t$ | 2.92643 | 0.43053 | 6.797 | 4.22e-09 | *** |

(residuals have autoregression) '***', '**', '*' and '.' describe significance levels at 0.001, 0.01, 0.05, and 0.1 respectively.

**SM Table 4.** Coefficients Final Cross-Correlation Model for Social Distancing (Daily Proportion)

|  | $Estimate$ | $Std.Error$ | $t.value$ | $Pr(>|t|)$ | Significance |
| --- | --- | --- | --- | --- | --- |
| $ar1$ | 0.4294 | 0.1480 | 2.9014 | 0.0051 | ** |
| $ma1$ | 0.7636 | 0.1004 | 7.6044 | 0.0000 | *** |
| $intercept$ | -60.2682 | 26.7714 | -2.2512 | 0.0279 | * |
| $c_{dist,t}$ | 0.1783 | 0.0261 | 6.8216 | 0.0000 | *** |
| $c_{dist,t-3}$ | -0.0452 | 0.0314 | -1.4366 | 0.1558 |  |
| $c_{dist,t-5}$ | 0.0941 | 0.0331 | 2.8410 | 0.0061 | ** |
| $c_{dist,t-6}$ | 0.0583 | 0.0305 | 1.9135 | 0.0603 | . |
| $c_{dist,t-8}$ | 0.0364 | 0.0313 | 1.1624 | 0.2495 |  |
| $c_{dist,t-11}$ | -0.0717 | 0.0307 | -2.3342 | 0.0228 | * |
| $c_{dist,t-12}$ | -0.0644 | 0.0276 | -2.3313 | 0.0230 | * |
| $c_{dist,t+2}$ | -0.1365 | 0.0213 | -6.4083 | 0.0000 | *** |
| $c_{dist,t+5}$ | -0.0506 | 0.0153 | -3.3149 | 0.0015 | ** |
| $t$ | 2.9318 | 0.6594 | 4.4460 | 0.0000 | *** |

'***', '**', '*' and '.' describe significance levels at 0.001, 0.01, 0.05, and 0.1 respectively.

**SM Table 5.** Coefficients of Final Cross-Correlation Model for Social Distancing (Daily Percentage)

|  | $Estimate$ | $Std.Error$ | $t\ value$ | $Pr(>|t|)$ | Significance |
| --- | --- | --- | --- | --- | --- |
| (Intercept) | 0.4942683 | 0.0854641 | 5.783 | 5.94e-08 | *** |
| $cp_{dist,t-1}$ | -0.4990231 | 0.0844211 | -5.911 | 3.27e-08 | *** |
| $t$ | 0.0008941 | 0.0001573 | 5.685 | 9.34e-08 | *** |

'***', '**', '*' and '.' describe significance levels at 0.001, 0.01, 0.05, and 0.1 respectively.

**ACF and pACF Tests for Model Validation**

**SM Figure 3.** ACF and pACF plots of the daily number of tweets for credible and misinformation relevant to wearing masks.

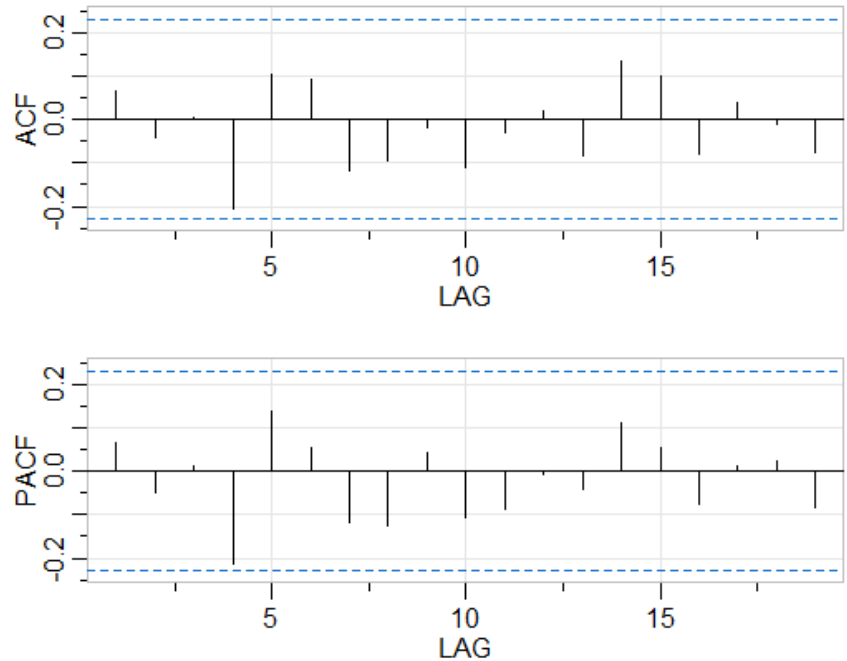

Note: The plots for residuals show that the autocorrelations at various lag times are all within the boundaries around 0, indicating no correlation structures. Thus, the final fitted model is valid.

**SM Figure 4.** ACF and pACF plots of the daily proportion of tweets for credible and misinformation relevant to wearing masks.

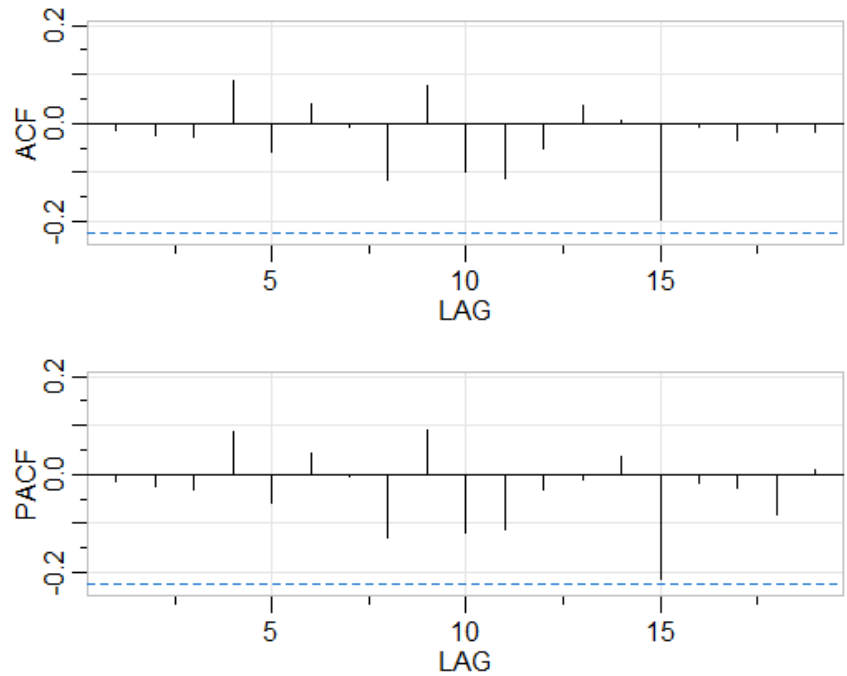

Note: Both ACF and pACF plots for residuals show that the autocorrelations at various lag times are all within the boundaries around 0, indicating no correlation structures. Thus, the final fitted model is valid.

**SM Figure 5.** ACF and pACF plots of the daily number of tweets for credible and misinformation relevant to social distancing.

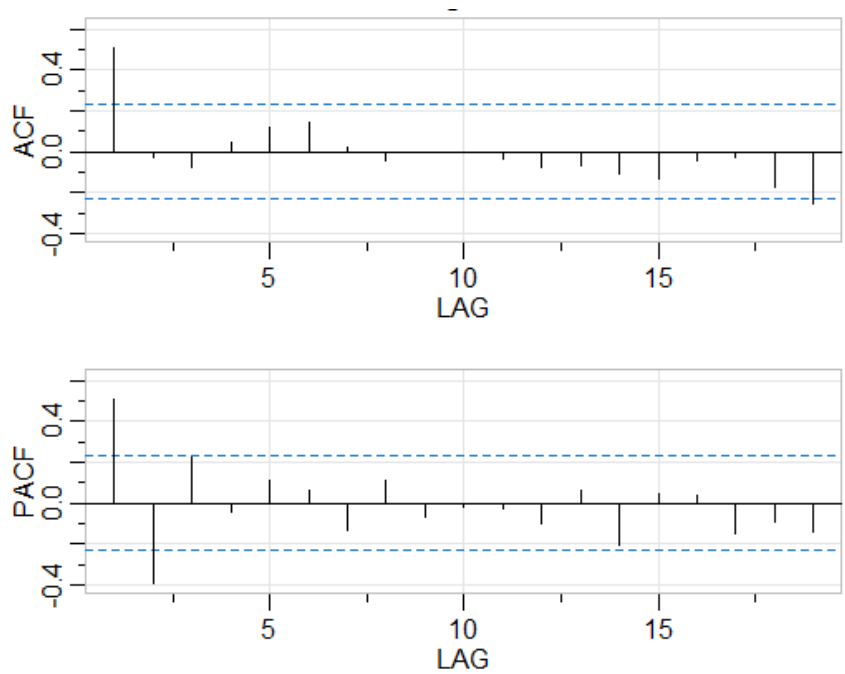

Note: The plots for residuals have peak values beyond the boundaries, indicating that the residuals have autocorrelation structures. Thus, we refit our model considering the time series structure of the residuals simultaneously.

**SM Figure 6.** ACF and pACF plots of the daily proportion of tweets for credible and misinformation relevant to social distancing.

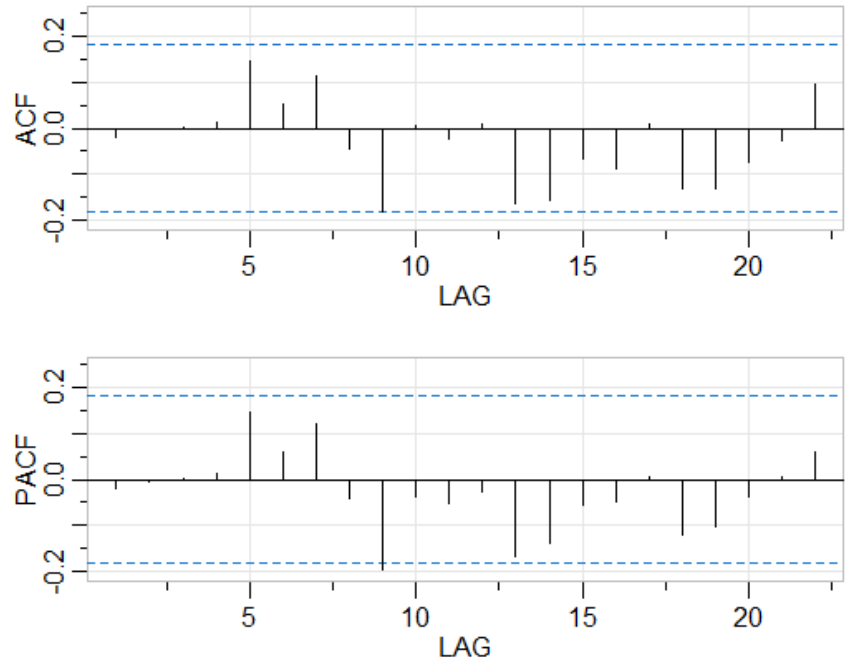

Both ACF and pACF plots for residuals show that the autocorrelations at various lag times are roughly within the boundaries around 0, indicating no correlation structures. Thus, the final fitted model is valid.